\documentclass[a4paper,fleqn,usenatbib]{mnras}

\usepackage[T1]{fontenc}
\usepackage{ae,aecompl}

\usepackage{graphicx}	
\usepackage{amsmath}	
\usepackage{amssymb}	

\usepackage{natbib}

\title[Modified gravity (MOG) accretion]{Black hole accretion in scalar--tensor--vector gravity}

\author[A. J. John]{
Anslyn J. John,$^{1}$\thanks{E-mail: a.john@ru.ac.za }
\\
$^{1}$Department of Mathematics, Rhodes University, Grahamstown 6139, South Africa\\
}

\date{Accepted XXX. Received YYY; in original form ZZZ}

\pubyear{2019}

\begin{document}
\label{firstpage}
\pagerange{\pageref{firstpage}--\pageref{lastpage}}
\maketitle

\begin{abstract}
We examine the accretion of matter onto a black hole in scalar--tensor--vector gravity (STVG) also known as modified gravity (MOG). The gravitational constant is $G=G_{N} (1 + \alpha)$ where $\alpha$ is a parameter taken to be constant for static black holes in the theory. The MOG black hole is spherically symmetric and characterised by two event horizons. The matter falling into the black hole obeys the polytrope equation of state and passes through two critical points before entering the outer horizon. We obtain analytical expressions for the mass accretion rate as well as for the outer critical point, critical velocity and critical sound speed. Our results complement existing strong field tests like lensing and orbital motion and could be used in conjunction to determine observational constraints on MOG.
\end{abstract}




\section{Introduction}

The concordance model of cosmology ($\Lambda$CDM) is a remarkably successful paradigm for the origin and development of large scale structure \citep{dodelson2003modern,weinberg2008cosmology}. In order to account for observed galaxy rotation curves, weak lensing and the formation of galaxy clusters the model requires most of the matter in the universe to only interact gravitationally. Cold dark matter necessitates an extension to the standard model of particle physics. The late time acceleration of the universe can be explained by introducing an energy component with a negative equation of state parameter viz. dark energy. The cosmological constant is a leading candidate for dark energy whilst other proposals include dynamical scalar fields e.g. quintessence.

An alternative school of thought to $\Lambda$CDM is to modify general relativity without introducing dark matter and dark energy. Hypothesised theories include TeVeS \citep{bekenstein2004} and $f(R)$ gravity \citep{nojiri2006}.

In the relativistic theory of gravitation known as scalar--tensor--vector gravity (STVG) or modified gravity (MOG) \citep{moffat2006scalar}, the gravitational `constant', $G$, as well as a vector field coupling, $\omega$, and the vector field mass, $\mu$, are treated as dynamical scalar fields. 
MOG is similar to earlier proposed modifications of general relativity viz. non--symmetric gravitational theory (NGT) \citep{moffat1995new} and metric--skew--tensor gravity (MSTG) \citep{moffat2005gravitational} in that they all introduce an extra degree of freedom due to skew--symmetric fields coupling to matter. In MSTG, which is the weak field approximation to NGT, the skew field is a rank three tensor, whilst in MOG the extra degrees of freedom are the scalar field $G(x)$, a rank one vector field and an effective scalar field mass associated with the vector field.

The trajectory of test particles in MOG obeys a modified acceleration law that provides a good fit to galaxy rotation curves \citep{brownstein2006galaxy,moffat2015rotational,
moffat2018ngc,de2018galaxy,green2019modified} and cluster data \citep{brownstein2006cluster,israel2018train} without invoking non--baryonic dark matter. The modified acceleration law adds a repulsive Yukawa force to the Newtonian law. This arises due to the exchange of a massive spin 1 boson whose effective mass and coupling to matter can vary with distance scale. Adding a scalar component to the Newtonian force law corresponds to exchanging a spin 0 particle and an attractive Yukawa force. Consequently a purely scalar correction cannot provide an acceleration law satisfying galaxy rotation curves and cluster data. Note that the MOG, MSTG, and NGT theories all satisfy cosmological tests. The degeneracy between these modified theories could be broken by testing their predictions in strong gravitational fields. This motivates our study of accretion onto MOG black holes. 

The final stage of gravitational collapse of a compact object in MOG is a static, spherically symmetric black hole \citep{moffat2015black}. This object has an enhanced gravitational constant, $G = G_N (1 + \alpha)$, and a repulsive gravitational force with charge $Q = \sqrt{\alpha G_N } M$ where $G_N$  is Newton's constant, $\alpha$ is a dimensionless parameter and $M$ is the black hole's mass. This black hole spacetime admits two event horizons and its Kruskal--Szekeres completion has been determined \citep{moffat2015black}.
In the same article, Moffat obtained a rotating black hole solution characterised by its mass, $M$, angular momentum, $a$, and the MOG parameter, $\alpha$. He also determined the motion and stability of a test particle in orbit about the black hole, the radius of the photon sphere and constructed a traversable wormhole solution. Strong and weak lensing by clusters \citep{moffat2018applying} and supermassive black holes \citep{izmailov2018modified} has been studied.
An earlier paper \citep{moffat2015modified} determined the sizes and shapes of shadows cast by MOG black holes. The possibility of observing MOG black holes with the Event Horizon Telescope has been explored \citep{guo2018observational,Moffat:2019uxp}. The dynamics of particles around rotating black holes \citep{sharif2017particle} and black holes immersed in magnetic fields has been investigated\citep{hussain}. The black hole quasinormal modes have been predicted\citep{manfredi2018quasinormal}, and the phenomena of superradiance\citep{wondrak2018superradiance} and weak cosmic censorship\citep{liang2019weak} have also been studied. The ADM mass and upper bounds on the spin of rotating MOG black holes have been determined\citep{sheoran2018mass}.

The gravitational wave events GW150914 \citep{abbott2016observation} and GW151226 \citep{abbott2016gw151226} observed by Advanced LIGO and VIRGO are believed to arise from the merger of binary black holes with masses in excess of 10 $M_{\odot}$. We currently lack a progenitor mechanism for black holes in this intermediate mass range. 
By appealing to MOG, the gravitational wave data can be reconciled with the merger of substantially less massive($M \leq 10 M_{\odot}$) black holes \citep{moffat2016ligo}. 
The dependence of merger rates of binary black holes on the enhanced gravitational constant has been calculated \citep{wei2018merger}.

A substantial body of literature in astrophysics is devoted to the problem of matter accreting onto stars and black  
holes \citep[e.g.][]{shapiro2008black, frank2002accretion, chakrabarti1990theory, shu1991physics}.
In the context of general relativity black hole accretion was studied by Michel \citep{michel1972accretion}. Shapiro determined the luminosity and frequency spectrum of gas accreting onto a black hole \citep{shapiro1973accretion} as well as the effects of an interstellar magnetic field \citep{shapiro1973magnetic}. He also solved the accretion problem for a rotating black hole \citep{shapiro1974accretion}. The significance of the gas backreaction on the accretion rate was explored by \citet{malec1999} and \citet{karkowski2006}. Charged black hole accretion was investigated by \citet{michel1972accretion} and \citet{ficek2015bondi} who included the effects of the cosmological constant. Accretion onto a broad class of static, spherically symmetric spacetimes was analysed by \citet{chaverra2015radial} and by \citet{bahamonde}. Studies of higher dimensional accretion were undertaken \citep{giddings2008astrophysical,john2013accretion} while quantum gravity corrections were also investigated \citep{yang}. Accretion in teleparallel gravity has been studied by \citet{ahmed}. A generalization of an earlier version \citep{john2016black} of this paper studied accretion disks around MOG black holes \citep{perez2017accretion}. 

In section~\ref{aaa} we briefly describe the action for modified gravity, outline the derivation of its static black hole solution and highlight its key features. In section~\ref{bbb} we determine the accretion rate for matter falling into the black hole. In section~\ref{ccc} we analyse the accretion rate and critical radius for various values of the modified gravity parameter, $\alpha$, and adiabatic index, $\gamma$. We state our conclusions in section~\ref{ddd}.

\section{Black holes in modified gravity}\label{aaa}

Modified gravity \citep{moffat2006scalar} belongs to the class of theories with varying fundamental constants. In addition to a modified Einstein--Hilbert action the theory introduces three scalar fields and a vector field. The action governing the theory is given by 
\begin{equation}
S = S_{grav} + S_{\phi} + S_{scalar} + S_{matter} \label{action}
\end{equation}
where
\begin{subequations} 
\begin{align}
S_{grav} & = \frac{1}{16 \pi} \int d^{4}x \sqrt{-g} \left[ \frac{1}{G} \left( R + 2 \Lambda \right) \right] \\
S_{\phi} & =  \int d^{4}x \sqrt{-g} \left[ \omega \left( \frac{1}{4} B^{ab} B_{ab} + V(\phi) \right) \right] \\
S_{scalar} & =  \int d^{4}x \sqrt{-g} \left[ \frac{1}{G^3} \left( \frac{1}{2} g^{ab} \nabla_a G \nabla_b G - V(G)\right) \right. \nonumber \\
& +  \frac{1}{G} \left( \frac{1}{2} g^{ab} \nabla_a \omega \nabla_b \omega - V(\omega) \right) \nonumber \\ 
& + \frac{1}{\mu^2 G} \left. \left( \frac{1}{2} g^{ab} \nabla_a \mu \nabla_b \mu - V(\mu) \right) \right]  
\end{align}
\end{subequations}
and $S_{matter}$ represents the action for the matter component.
$S_{grav}$ is the standard Einstein--Hilbert action where Newton's constant $G_{N}$ has been promoted to a dynamical scalar field, $G(x^{a})$. In $S_{\phi}$ we have a Maxwell--like contribution to the action from the phion vector field $\phi_{a}$ which is defined via 
\begin{equation}
B_{ab} = \partial_a \phi_b - \partial_b \phi_a. \label{phion}
\end{equation}
Each of the three scalar fields viz. $G$, $\omega$ and $\mu$ has an associated potential, $V$. 

The static, spherically symmetric black hole solution for MOG was obtained \citep{moffat2015black} by solving the vacuum field equations derived from the action (\ref{action}). The matter energy--momentum tensor vanishes ($T_{matter} = 0$) and we neglect the influence of the cosmological constant ($\Lambda = 0$). The enhanced gravitational coupling, $G = G_{N}(1 + \alpha)$, is taken to be constant i.e. $\partial_{a} G = 0$. The field coupling the vector field, $\phi_{a}$, to the action is also taken to be constant viz. $\omega = 1$. The energy--momentum tensor due to the vector field is given by $T_{ab}^{(\phi)} = -\frac{1}{4 \pi} \left( B_{b}^{\,c} B_{ac} - \frac{1}{4} B^{ab} B_{ab} \right)$. In order to successfully reproduce galaxy rotation curves and cluster dynamics the vector field mass has to be $m_{\phi} = 2.6 \times 10^{-28} \mathrm{eV}$, which corresponds to a scale of $0.042 (\mathrm{kpc})^{-1}$. The field mass is negligible on the scale of compact objects and can be safely ignored for black holes in the theory. The vacuum field equations for the phion vector field are given by 
\begin{subequations}
\begin{eqnarray}
\nabla_{b} B^{ab} &=& 0 \\
\nabla_{c} B_{ab}  + \nabla_{b} B_{ca} + \nabla_{a} B_{bc} &=& 0.
\end{eqnarray}
\end{subequations}
The spherically symmetric spacetime due to a black hole of mass $M$ in MOG is described by the line element
\begin{equation}
\begin{split}
ds^2 &= \left( 1 - \frac{2GM}{r} + \frac{\alpha G G_N M^2}{r^2}  \right) dt^2 \\
&- \left( 1 - \frac{2GM}{r} + \frac{\alpha G G_N  M^2}{r^2}  \right)^{-1} d r^2 \\
& - r^2 d \theta^2 - r^2 \sin^2 \theta d \phi^2 \label{metric}
\end{split}
\end{equation}
where $G = G_N (1 + \alpha)$, $G_N$ is Newton's gravitational constant and $\alpha$ is a dimensionless constant. In the complete MOG theory the fundamental `constants' vary with time. The black hole solution (\ref{metric}) however is static; hence the modified gravitational constant is implicitly fixed. The speed of light is normalised. The Schwarzschild solution of general relativity is recovered in the limit where $\alpha \rightarrow 0$. 

The spacetime (\ref{metric}) is asymptotically flat and singular at the origin. It admits two event horizons viz. $r_{\pm} = G_N M \left[ 1 + \alpha \pm \left( 1 + \alpha \right)^{1/2} \right]$ and is formally similar to the Reissner-Nordstr\"{o}m line element describing a charged black hole. As in that case the inner horizon is a Cauchy horizon which we expect to also be unstable.

The similarity of the MOG black hole metric to the Reissner-Nordstr\"{o}m solution is unsurprising given the presence of a Maxwell--like vector field. The MOG vector field is a $4$--potential sourced by the gravitational `charge' viz. mass. In Einstein--Maxwell theory the analogous potential is sourced by the electric charge. Electrically charged black holes have little astrophysical significance as they are ephemeral. Any hypothetical charged compact object will accrete charges of the opposite sign and rapidly neutralize \citep{wald1974black}. The gravitational charge in MOG only has one sign and the MOG black hole will not be short--lived. 

\section{Accretion in MOG}\label{bbb}

\subsection{Conservation laws}

In spherically symmetric accretion the gas surrounding a non--rotating black hole is initially at rest. Under the influence of the black hole's gravitational attraction the gas accelerates inwards. The gas velocity reaches its local sound speed and then continues to accelerate towards the black hole at supersonic velocities.

The gas accreting onto the black hole is modelled as a perfect fluid with energy--momentum tensor 
\begin{equation}
T^{ab} = \left( \rho + p \right) u^a u^b - p g^{ab}	\label{stress}
\end{equation}
where $\rho$, $p$ and $u^a$ are the fluid's energy density, pressure and $4$--velocity respectively. Since the gas flow is stationary and spherically symmetric, its only non--vanishing velocity components are 
$u^{0}(r)$ and $u^{1} \equiv v(r)$.
Under the normalisation condition $u^a u_a = 1$ the temporal component of the $4$--velocity is 
\begin{equation}
u^{0} = \frac{\sqrt{1 - \frac{2GM}{r} + \frac{\alpha G_N G M^2}{r^2}  + v^2} }{1 - \frac{2GM}{r} + \frac{\alpha G_N G M^2}{r^2}  }.
\end{equation}
Note that we have implicitly neglected the phion field \eqref{phion} of the accreting fluid by assuming it to be a subdominant contribution to the total stress tensor. The fluid will, in general, possess a phion charge that contributes to the full stress tensor in a manner analogous to the Faraday tensor in electromagnetism. For sufficiently low values of $\alpha$ and low fluid densities, $\rho$, this should be a reasonable approximation. A more complete analysis should incorporate this feature\footnote{I thank Martin Green for highlighting the significance of the phion field.}.

If particle number is conserved during the flow then 
\begin{equation}
\nabla_a \left( n u^a \right) = 0 \label{particle}
\end{equation}
where $n$ is the fluid's number density and $\nabla_a$ is the covariant derivative with respect to the coordinate $x^a$.
Conservation of energy--momentum is governed by
\begin{equation}
\nabla_a T^{a}_{\: \: b} = 0. \label{energy-mom}
\end{equation}
For a perfect fluid accreting onto the black hole (\ref{metric}), the continuity equation (\ref{particle}) is 
\begin{equation}
\frac{1}{r^2} \frac{d}{d r} \left( r^2 n v \right) = 0 \label{numcon}
\end{equation}
while equation (\ref{energy-mom}) can be re--written as 
\begin{align}
& 0  =  \frac{1}{r^2} \frac{d}{d r} \left[ r^2 \left( \rho + p \right) v \left( 1 - \frac{2GM}{r} + \frac{\alpha G_N G M^2}{r^2} + v^2 \right)^{1/2} \right] \label{encon} \\
& v \frac{d v}{d r}  = - \frac{d p}{d r} \frac{\left( 1 - \frac{2GM}{r} + \frac{\alpha G_N G M^2}{r^2} + v^2 \right)}{\rho + p} \nonumber \\
& - \frac{GM}{r^2} \left( 1 - \frac{\alpha G_N M}{r}  \right). \label{tov}
\end{align}
We restrict our attention to adiabatic flows so the first law of thermodynamics for the fluid is given by 
\begin{equation}
T ds = 0 = d \left( \frac{\rho}{n} \right) + p d \left( \frac{1}{n} \right) \label{energy}
\end{equation}
which, upon integration, yields
\begin{equation}
\frac{d \rho}{d n} = \frac{\rho + p}{n}.	\label{firstlaw}
\end{equation}
Using the fluid's adiabatic sound speed, 
\begin{equation}
a^2 \equiv \frac{d p}{d \rho} = \frac{d p}{d n} \frac{n}{\rho + p}, \label{sound}
\end{equation}
we express the continuity (\ref{numcon}) and momentum (\ref{tov}) equations as
\begin{align}
&\frac{v'}{v} +  \frac{n'}{n} = -\frac{2}{r} \label{dynone} \\
&v v' + \left( 1 - \frac{2GM}{r} + \frac{\alpha G G_N M^2}{r^2} + v^2 \right) a^2 \frac{n'}{n} \nonumber \\
& =  - \frac{GM}{r^2} \left( 1 - \frac{\alpha G_N M}{r} \right) \label{dyntwo}
\end{align}
where primes denote spatial derivatives.
The number density and velocity derivatives can be written as
\begin{align}
n' &= \frac{D_1}{D} \label{crit-one} \\
v' &= \frac{D_2}{D} \label{crit-two}
\end{align}
where we have defined 
\begin{align}
D & = \frac{v^2 - \left( 1 - \frac{2GM}{r} + \frac{\alpha G G_N M^2}{r^2} + v^2 \right)a^2 }{nv} \\
D_1 & = - \frac{1}{v} \left( \frac{2 v^2}{r} - \frac{GM}{r^2} \left( 1 - \frac{ \alpha G_N M}{r} \right) \right) \\
D_2 & = \frac{1}{n} \left[ \left( 1 - \frac{2GM}{r} + \frac{\alpha G G_N M^2}{r^2} + v^2 \right) \frac{2 a^2}{r} \right. \nonumber \\
& - \left. \frac{GM}{r^2} \left( 1 -  \frac{\alpha G_N M}{r}  \right)\right].
\end{align}
Introducing $m$, the mass of an individual gas particle, we obtain the mass accretion rate by integrating the continuity equation (\ref{numcon}) over a unit volume
\begin{equation}
\dot{M} = 4 \pi m n r^{2} v. \label{mdot}
\end{equation}
The accretion rate, $\dot{M}$ has dimensions of $\mathrm{mass}.\mathrm{time}^{-1}$ and is independent of $r$.
Equations (\ref{numcon}) and (\ref{encon}) can be combined to yield 
\begin{equation}
\left( \frac{\rho + p}{n} \right)^2 \left( 1 - \frac{2GM}{r} + \frac{\alpha G G_N  M^2}{r^2} + v^2 \right) = E \label{bernone}
\end{equation}
which is the relativistic version of the Bernoulli equation. If the gas is at rest at large distances from the black hole i.e. $v_{\infty} = 0$ the integration constant is $E = \left(\frac{\rho_{\infty} + p_{\infty}}{n_{\infty}}\right)^2$ and has dimensions of enthalpy squared.

\subsection{Critical points}

The gas is at rest very far from the black hole. Under the influence of the black hole's gravitational field it accelerates inwards, eventually falling into the outermost event horizon. In the original Bondi problem (accretion driven by a Newtonian potential) the gas accelerates from rest and passes through a critical point, where its velocity matches its local sound speed. The gas then flows towards the central mass at supersonic velocities in a manner analogous to flow through a de Laval nozzle \citep{shu1991physics}. A similar velocity profile occurs for accretion onto a Schwarzschild black hole. In this case, however, the gas velocity at the critical point does not equal its sound speed. Here we establish the existence of two critical points in accretion onto a MOG black hole.

A critical point occurs whenever the quantity $D$ in (\ref{crit-one}) - (\ref{crit-two}) vanishes. In order to avoid infinite acceleration the expressions $D_1$ and $D_2$ must simultaneously vanish. The critical point conditions are thus $D= D_1 = D_2 = 0$ at particular values of $r$.

At large distances the gas velocity is subsonic i.e. $v < a$. Moreover for conventional matter the sound speed is always subluminal i.e. $a \leq 1$. Thus we have $D \approx \frac{v^2 - a^2}{nv}$ for large values of $r$. Since the gas flows inwards we have $v < 0$ and thus $D>0$. At the outermost horizon we have $D = \frac{v}{n} (1 - a^2)$ hence $D<0$. Thus $D=0$ at some distance $r_s$ between the outer event horizon and infinity i.e. there is at least one critical point satisfying $r_{H}^{+} < r_s < \infty$.

The critical point conditions viz. $D = D_1 = D_2 = 0$ at $r_s$ are 
\begin{align}
& v^{2}_{s}  - \left( 1 - \frac{2GM}{r_{s}} + \frac{\alpha G G_N M^2}{r^{2}_{s}} + v^{2}_{s} \right)a^{2}_{s} =  0 \\
& \left( 1  - \frac{2GM}{r_{s}} + \frac{\alpha G G_N M^2}{r^{2}_{s}} + v^{2}_{s} \right) \frac{2 a^{2}_{s}}{r_{s}} \nonumber \\
& - \frac{GM}{r^{2}_{s}} \left( 1 - \frac{ \alpha G_N M}{r_{s}} \right) =  0  \\
& \frac{2 v^{2}_{s}}{r_{s}}  - \frac{GM}{r^{2}_{s}} \left( 1 -  \frac{ \alpha G_N M}{r_{s}} \right) =  0
\end{align}
where $v_s \equiv v(r_{s})$ etc.
Introducing the dimensionless variable $y \equiv \frac{GM}{r}$ the critical points are located at 
\begin{equation}
y_{s}^{\pm} = \frac{3 a^{2}_{s} + 1}{ \frac{2 \alpha}{\alpha + 1} a^{2}_{s} + 1} \left[ 1 \pm \sqrt{\Delta} \right]
\end{equation}
where 
\begin{equation}
\Delta \equiv 1 - \frac{ \frac{ 8 \alpha}{\alpha + 1} (a^{2}_{s} + 1) a^{2}_{s}}{(3 a^{2}_{s} + 1)^2}.
\end{equation}
We confine our attention to the outermost critical point which we label
\begin{equation}
y_s = \frac{3 a^{2}_{s} + 1}{ \frac{ 2 \alpha}{\alpha + 1} a^{2}_{s} + 1} \left[ 1 + \sqrt{\Delta} \right].
\end{equation}
At this point the critical velocity is given by 
\begin{eqnarray}
v_{s}^{2} &=& \frac{1}{2}y_{s} \left( 1 - \frac{\alpha}{\alpha + 1 } y_{s} \right) \\
&=& \frac{(3 a^{2}_{s} + 1) (a^{2}_{s} + 1) (1 - \sqrt{\Delta})}{ \frac{ 4 \alpha}{\alpha + 1} (a^{2}_{s} + 1)^2 }.
\end{eqnarray}

\subsection{The accretion rate}

We now evaluate the Bernoulli equation (\ref{bernone}) at the critical point to determine the critical sound speed, $a_s$, in terms of $a_{\infty}$. We employ a polytropic equation of state for the gas viz.
\begin{equation}
p = K n^{\gamma} \label{poly}
\end{equation}
where $1 < \gamma < 5/3$. For this polytrope the energy equation (\ref{energy}) can be integrated to obtain
\begin{equation}
\rho = \frac{K}{\gamma - 1}n^{\gamma} + mn \label{rho_int}
\end{equation}
where $mn$ is the rest--energy density.
Using (\ref{sound}) the Bernoulli equation (\ref{bernone}) is rewritten as
\begin{align}
& \left( 1 - \frac{1}{\gamma - 1} a^2 \right)^2 \left( 1 - \frac{2GM}{r} + \frac{\alpha G G_N M^2}{r^2} + v^2 \right)^{-1} \nonumber \\
& = \left( 1 - \frac{1}{\gamma - 1} a^{2}_{\infty} \right)^2. \label{bernoulli}
\end{align}
We use the fact that $a \leq a_s \ll 1$ when the gas is still in the non--relativistic regime \citep{john2013accretion} to evaluate (\ref{bernoulli}) at the critical point, $y_s$, and obtain
\begin{equation}
(1 + 3 a^{2}_{s}) \left( 1 - \frac{2}{\gamma - 1 } a^{2}_{s}  \right) \approx 1 - \frac{2}{\gamma - 1 } a^{2}_{\infty}
\end{equation}
The critical sound speed, to leading order, is thus 
\begin{equation}
a_{s}^{2} = \frac{2}{5 - 3 \gamma } a^{2}_{\infty}.
\end{equation}
We can determine the critical number density, $n_s$, by combining (\ref{sound}), (\ref{poly}) and (\ref{rho_int}) to obtain 
\begin{eqnarray}
\gamma K n^{\gamma - 1} &=& \frac{m a^{2}}{1 - a^{2}/ \left( \gamma - 1 \right) } \\
&\approx& m a^2 
\end{eqnarray}
where we have exploited the relation $a^2 / \left( \gamma - 1 \right) \ll 1$.
Since $n \sim a^{2/(\gamma - 1)}$ we have 
\begin{equation}
\left( \frac{n_s}{n_{\infty}} \right) \approx \left( \frac{a_s}{a_{\infty}} \right)^{\frac{2}{\gamma - 1}}. \label{numden}
\end{equation}
The mass accretion rate, $\dot{M}$, is independent of $r$. In particular, equation (\ref{mdot}) must hold at the outer critical point, $r_s$, hence
\begin{equation}
\dot{M} = 4 \pi m n_s r^{2}_{s} v_{s}.
\end{equation}
The rate at which polytropic matter accretes adiabatically onto a MOG black hole is  
\begin{align}
& \dot{M} = \pi \left( GM \right)^2   \frac{m n_{\infty}}{a_{\infty} } \left( \frac{5 - 3 \gamma}{2} \right)^{\frac{\gamma - 3}{2 (\gamma -1)} } \left[ 1 + \frac{3 a_{\infty}^{2} }{ (5 - 3 \gamma ) } \right] \nonumber \\
& \times \left[  \left( \frac{5 - 3 \gamma}{a_{\infty}} \right)^2 + 6 -  \frac{ 2 \alpha}{\alpha + 1} \right]. \label{mdotstvg}
\end{align}

\section{Analysis}\label{ccc}

In the limit $\alpha \rightarrow 0$ the accretion rate (\ref{mdotstvg}) reduces to the formula for accretion onto a Schwarzschild black hole \citep{michel1972accretion} viz.
\begin{equation}
\dot{M} = 4 \pi \left( \frac{GM}{a_{\infty}^{2}} \right)^2 m n_{\infty} a_{\infty} \left( \frac{1}{2} \right)^{\frac{\gamma + 1}{2 (\gamma -1)}} \left( \frac{5 - 3 \gamma}{4} \right)^{\frac{3 \gamma - 5}{2 (\gamma - 1)}}.  \label{michel} 
\end{equation}
Note that as with spherical accretion in general relativity, the accretion rate onto a MOG black hole is proportional to the square of the black hole's mass i.e. $\dot{M} \propto M^2$. We will only consider solar mass black holes i.e. $M = M_{\odot} = 1.9884 \times 10^{33} \mathrm{g}$. The accreting gas is taken to be ionized hydrogen with molecular mass $m_H = 1.6727 \times 10^{-24} \mathrm{g}$ at temperature $T = 10^4 \mathrm{K}$ and number density $n_{\infty} = 1 \mathrm{cm}^{-3}$. For an ideal gas the sound speed is $a^2 = \gamma k_B T / \mu m$. The mean molecular weight for ionized hydrogen is $\mu = 1/2$. We fix the adiabatic index, $\gamma = 5/3$, of the gas at the boundary since this is typical for ionized hydrogen in stellar environments. We do however consider a range of values of $\gamma$ as the gas falls towards the black hole. The boundary condition for the gas sound speed is thus $a_{\infty}^{2} = 2.7513 \times 10^{12} \mathrm{cm}^2 \mathrm{s}^{-2}$. For our analysis we utilize geometric units where $G = c = 1$.

\begin{figure}
\includegraphics[width=\columnwidth]{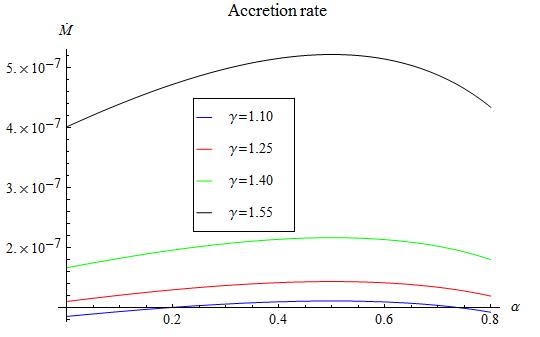}

\caption{\label{figone} (color online) The accretion rate, $\dot{M}$, as a function of the MOG parameter, $\alpha$, for various adiabatic indices, $\gamma$. Values are expressed in geometric units.}
\end{figure}

In Fig. \ref{figone} we plot the mass accretion rate, $\dot{M}$, as a function of the gravitational parameter, $\alpha$. We do not consider very large values of $\alpha$ as this would imply significant deviations away from the standard value of Newton's constant. The values for $\dot{M}$ in general relativity, where a polytrope accretes onto a Schwarzschild black hole are recovered at $\alpha = 0$. Here the accretion rate increases as the adiabatic index increases. A gas that is close to the isothermal limit, $\gamma = 1$, accretes at a lower rate than relativistic gases, $\gamma = 4/3$. Non--relativistic gases, $\gamma = 5/3$, accrete at the fastest rate. 

This behaviour persists when one looks at MOG accretion. A non--relativistic gas accretes at a greater rate than a relativistic gas or an isothermal gas. The accretion rate for each class of gas rises gently then slowly decreases as the MOG parameter, $\alpha$ is increased. The most pronounced increase occurs as we approach the non--relativistic limit, $\gamma = 5/3$. For gases with lower adiabatic indices the change in accretion rate is quite small. Even for the highly idealised case of spherical, adiabatic accretion it appears to be difficult to distinguish MOG from general relativity. 

Spherical accretion is typically an inefficient process for converting gravitational energy into radiation. Rotating black holes accreting non--adiabatic gases are suspected to be responsible for the energy emitted by active galactic nuclei \citep{frank2002accretion}. In this case the efficiency of energy conversion is significantly higher than for spherical accretion.

\section{Conclusion}\label{ddd}

The scalar--tensor--vector theory of gravity (STVG), or modified gravity (MOG), satisfies a number of cosmological tests. The black hole solutions of the theory possess two event horizons and depend on the black hole's mass, $M$, angular momentum, $a$, and the MOG parameter, $\alpha$, which characterises deviations from the gravitational constant, $G$.

We studied the accretion of a polytropic gas onto a non--rotating black hole in MOG. The gas is at rest far from the black hole, then accelerates towards its outer event horizon. We established the existence of a critical point, $r_s$ in the flow and calculated its location as well as the gas velocity, $v_s$ and sound speed, $a_s$, at the critical point. We determined an analytical expression for the rate at which gas accretes onto the black hole. The accretion rate, $\dot{M}$ is parametrised by $\alpha$ and the adiabatic index, $\gamma$. In the limit that $\alpha \rightarrow 0$ we recover the accretion rate for matter falling into a Schwarzschild black hole. As with the Schwarzschild case gases with higher adiabatic indices accrete at a faster rate. As the MOG parameter, $\alpha$ increases, the accretion rate for the gas increases then decreases slightly. The gas properties, characterised by $\gamma$, have a greater effect on the accretion rate than the change in gravitational theory, parametrised by $\alpha$. Uncertainty in gas dynamics thus dominates uncertainty in the gravitational theory. Since changes in the accretion rate in this idealised adiabatic, spherical problem are quite subtle it would appear to be quite difficult to distinguish between general relativity and MOG using accreting systems alone. The phion charge of the accreting fluid should be included in a more complete analysis. Strong field tests of MOG should incorporate accretion dynamics as well as lensing and test particle motion.

\section*{Acknowledgements}

The author thanks Rhodes University for financial support. He also acknowledges useful discussions with Martin Green and John Moffat.




\bibliographystyle{mnras}
\bibliography{svt} 








\bsp	
\label{lastpage}
\end{document}